\documentclass[aps,prl,reprint,superscriptaddress,longbibliography,nofootinbib,natbib]{revtex4-1}
\usepackage{lipsum}

\usepackage[utf8]{inputenc}

\usepackage{amsfonts}
\usepackage{amsmath}
\usepackage{tikz}
\usetikzlibrary{arrows,calc}
\usetikzlibrary{patterns,snakes}

\usepackage{ upgreek }

\def\({\left(} 
\def\){\right)}
\def\[{\left[} 
\def\]{\right]}

\newcommand{\Z}{\ensuremath{\mathbb Z}}

\newcommand{\be}{\begin{equation}}
\newcommand{\ee}{\end{equation}}
\newcommand{\non}{\nonumber \\}
\newcommand{\reef}[1]{(\ref{#1})}
\renewcommand{\eqref}[1]{(\ref{#1})}

\newcommand{\eg}{{\it e.g.,}\ }

\newcommand{\bea}{\begin{eqnarray}}
\newcommand{\eea}{\end{eqnarray}}

\begin{document}


\title{A model of persistent breaking of discrete symmetry} 

\author{Noam Chai}
\affiliation{The Racah Institute of Physics, The Hebrew University of Jerusalem, \\ Jerusalem 91904, Israel \\[4pt]}
\author{Anatoly Dymarsky}
\affiliation{Department of Physics and Astronomy, \\ University of Kentucky, Lexington, KY 40506\\[4pt]}
\affiliation{Skolkovo Institute of Science and Technology, \\ Skolkovo Innovation Center, Moscow, Russia, 143026\\[2pt]}
\author{Michael Smolkin}
\affiliation{The Racah Institute of Physics, The Hebrew University of Jerusalem, \\ Jerusalem 91904, Israel \\[4pt]}

\date{\today}
\begin{abstract}
We show there exist  UV-complete field-theoretic models in general 
dimension, including $2+1$,  with the spontaneous  breaking of a global symmetry, which persists to the arbitrarily
 high temperatures. 
Our example is a conformal vector model with the $O(N)\times \mathbb{Z}_2$ symmetry at zero 
temperature. Using conformal perturbation theory we establish $\mathbb{Z}_2$ symmetry 
is broken at finite temperature for  
$N>17$. Similar to  recent constructions of \cite{Chai:2020zgq,Chai:2020onq}, in the infinite $N$ limit our model has a non-trivial conformal 
manifold, a moduli space of vacua, which gets deformed at finite temperature. Furthermore, in this regime the model 
admits a persistent breaking of $O(N)$ in $2+1$ dimensions, therefore providing another example where the 
Coleman-Hohenberg-Mermin-Wagner theorem can be bypassed.
\end{abstract}

\pacs{}

\maketitle

\section{Introduction}

The phenomenon of spontaneous symmetry breaking is ubiquitous: many real systems as well as field theoretic models exhibit spontaneous breaking of both discrete and continuous symmetries at zero or sufficiently small temperature. The conventional picture suggests the full symmetry is restored for sufficiently high temperatures. There are also situations when the low-temperature phase  is symmetric but symmetry is broken as the temperature increases. In this case too one normally expects the symmetry to be eventually restored for even higher temperatures, and there are seemingly many theoretical results supporting such a conclusion. Yet as we discuss below they rely on the stringent assumptions that can be evaded. Which raises the question -- is spontaneous breaking that persists to arbitrarily high temperatures possible? In this letter we answer this question by constructing UV-complete conformal field-theoretical models in diverse dimensions which exhibit persistent breaking of both discrete and continuous symmetries.  

For lattice systems it is well appreciated the symmetry is restored for temperates 
large in comparison with the lattice spacing \cite{bratteli2012operator}. This suggests in effective field theory spontaneous breaking is possible up to UV scale, as illustrated by the UV-incomplete example of \cite{Weinberg:1974hy}. 
Yet the lattice-based arguments are not applicable to  UV-complete field theoretic models, which we focus on. 
There is also famous  Coleman-Hohenberg-Mermin-Wagner theorem \cite{Mermin:1966fe,PhysRev.158.383} and its generalizations, which rule out the possibility of the longer range order at $T\neq 0$ in a wide class of two-dimensional $d=2+1$ systems. This seemingly prohibit spontaneous breaking of  a continuous symmetry, but here too there are many assumptions and exceptions, starting from the  example of   \cite{PhysRev.176.250}. Importantly to what follows, the original work
 \cite{PhysRev.158.383} already notes the argument may break down if the lattice model exhibits long-range  interactions.  
 Thus, there is a phase transition   in a Heisenberg magnet with suitably adjusted long-range interactions, see \eg \cite{Halperin_2018} for a recent review on the Coleman-Hohenberg-Mermin-Wagner theorem and its limitations. 
This suggests non-local  field-theories, which may result from such lattice models in the continuous limit may be immune to various no-go results.

To better illustrate this idea we briefly mention Coleman's no-go theorem \cite{Coleman:1973ci}, which in $d=1+1$ excludes spontaneous symmetry breaking because  the corresponding Goldstone bosons, being massless, would have infrared divergences. This argument works for the  short-range interactions, whereas as above introduction of the long-range forces allows phase-transitions in the one dimensional systems \cite{Dyson:1968up,Dyson1969,Dyson1971,Laflorencie_2005,PhysRevLett.119.023001}. The no-go results in various dimensions are related and can be evaded simultaneously: our model exhibits  persistent breaking for $1<d<3+1$.

The discussion above mostly applies to continuous symmetries. For discrete symmetries
even less is known. As for continuous symmetries, there is no universal theoretical argument requiring discrete symmetry to be restored at high temperatures. At the same time to the best of our knowledge  there were  
no examples of persistent symmetry breaking in $2+1$ dimensions. 
In higher dimensions there are CFTs in $d=4-\epsilon$ \cite{Chai:2020zgq,Chai:2020onq,Chai:2020hnu}  and $d=4$   \cite{Chaudhuri:2020xxb,Chaudhuri:2021dsq}  which have some of their internal symmetries broken at arbitrary finite temperature. Yet these models are not free of criticism. In the former case $\epsilon$ can not be taken to one and therefore theories in question are not necessarily unitary \cite{Hogervorst:2015akt}. And the latter case of \cite{Chaudhuri:2020xxb} is inconclusive because of the possible impact of $1/N$ corrections.
We also refer the reader to \cite{Bajc:2021ope}, where asymptotically safe theories \cite{Litim:2014uca} in $d=3+1$ were  considered in the context of persistent symmetry breaking. Another idea  explored in the literature is placing the theory in a curved spacetime. Thus,  the $O(N)$ model in AdS evades the Coleman-Hohenberg-Mermin-Wagner theorem \cite{Carmi:2018qzm} but at high temperatures the symmetry is restored.
There are other AdS/CFT candidates of persistent order \cite{Buchel:2009ge,Donos:2011ut,Gursoy:2018umf,Buchel:2018bzp,Buchel:2020thm,Buchel:2020xdk,Buchel:2020jfs} which are perturbatively stable, but the symmetric phase has smaller free energy.\footnote{See \cite{Buchel:2021ead} where the holographic conformal order is studied on $S^3$.} Other non-unitary models of persisten breaking in the presence of  chemical potential include \cite{Hong:2000rk,Komargodski:2017dmc,Aitken:2017ayq,Tanizaki:2017qhf,Dunne:2018hog,Wan:2019oax}. We should also mention the renowned Berezinskii–Kosterlitz–Thouless  transition \cite{Berezinsky:1970fr,Berezinsky:1972rfj,Kosterlitz:1973xp}  as an example evading the Coleman-Hohenberg-Mermin-Wagner theorem yet not leading to persistent symmetry breaking. 

The main goal of this letter is to construct an example which would be free of aforementioned deficiencies. 
Our model is a conformal, and hence UV-complete vector theory  which breaks discrete symmetry at finite  temperature $T$. Because of scale invariance this breaking persists to arbitrary high $T$. The result holds true in $2+1$ and extends to other $1<d<3+1$ where the model is manifestly stable.\footnote{Stability suggests the model is unitary; it would be interesting to confirm this by calculating anomalous dimensions at the IR fixed point. We also expect the fixed point to be a CFT, but since our model is non-local strictly speaking this has to be verified along the lines \cite{Paulos:2015jfa}.}  Generalizations of our model exhibit persistent breaking of continuous symmetries in $d=2+1$ and beyond \cite{followup}. Our construction bypasses  the Coleman-Hohenberg-Mermin-Wagner theorem due to  its non-local nature. It can be viewed as two weakly interacting copies of the Long Range Ising (LRI) model \cite{Dyson:1968up,PhysRevLett.29.917,PhysRevB.8.281}.\footnote{For original and recent studies of the LRI model see   \cite{Honkonen_1989,LEGUILLOU1990559,El-Showk:2013nia,PhysRevLett.89.025703,Picco:2012ak,Blanchard:2012xv,El-Showk:2013nia,Angelini_2014,Paulos:2015jfa,Defenu_2015,Behan:2017dwr}.}

\section{The model}
To begin with, we introduce our model.
Consider the following Gaussian action in $1\leq d < 4$ dimensions,
\bea
\nonumber
 S_0= &\mathcal{N}_\phi& \int d^d x_1 \int d^d x_2 {\vec \phi(x_1)\cdot \vec\phi(x_2) \over |x_1-x_2|^{2(d-\Delta_\phi)}} + \\
  &\mathcal{N}_\sigma& \int d^d x_1 \int d^d x_2 {\sigma(x_1) \sigma(x_2) \over |x_1-x_2|^{2(d-\Delta_\sigma)}} ~. 
\label{action}
\eea
The fundamental fields include  scalars   $\vec \phi$ and $\sigma$ transforming in vector and singlet representations of $O(N)$. 
The model also admits a $\mathbb{Z}_2$ symmetry that flips the sign of $\sigma$.
The coefficients $\mathcal{N}_\phi$ and $\mathcal{N}_\sigma$ are fixed so that the two-point functions of $\vec\phi$ and $\sigma$ are canonically normalized. 
For brevity we suppress vector indices in what follows. The scaling dimensions of the generalized free fields are chosen to be
\be
 \Delta_\phi={d-\epsilon_1\over 4}\,,  ~ \Delta_\sigma={d-\epsilon_3\over 4},
\ee
with $\epsilon_{1,3}\ll 1$, so that the following quartic operators become weakly relevant 
\be
\label{Oi}
 \mathcal{O}_1= (\phi^2)^2~,\, \, \mathcal{O}_2= \phi^2 \sigma^2 ~,\, \,  \mathcal{O}_3= \sigma^4,
\ee
with scaling dimansions $\Delta_1=4\Delta_\phi$, $\Delta_2=2(\Delta_\phi+\Delta_\sigma)$ and $\Delta_3=4\Delta_\sigma$.  
This model is conformal, we list OPE coefficients $C_{ij}^k$ and other technical details in Supplemental Material. 

Next, we consider the following deformation 
\be
 S=S_0 + \sum_{i=1}^3 {g_i \mu^{\epsilon_i} \over N} \int d^dx \, \mathcal{O}_i(x)  ~,
 \label{Sdeform}
\ee
where $\epsilon_i\ll 1$.\footnote{Note that $\epsilon_2=(\epsilon_1+\epsilon_3)/2$.} It induces an RG flow of the form
\be
 \mu {dg_i\over d\mu} = -\epsilon_i g_i + {\pi^{d/2} \over N\Gamma\({d\over 2}\)} \sum_{j,k} C_{jk}^i g_j g_k + \ldots
 \label{RGflow}
\ee
As we argue below, for $N> 5$
there is a fixed point with $g_2<0$. Moreover, the IR CFTs with negative 
$g_2$ are stable. They define a class of theories with the persistent symmetry breaking. 

To understand the unbroken symmetries of the IR critical point at finite temperature 
we consider the effective potential, $V_\text{eff}$, for the zero mode.  To leading order in $\epsilon_i$, thermal 
fluctuations simply induce quadratic terms in addition to the quartic potential  \reef{Sdeform},
\bea
\nonumber
 V_\text{eff}(\phi,\sigma; \beta)= \mathcal{M}_\phi(\beta) \phi^2 + \mathcal{M}_\sigma(\beta) \sigma^2 
 +  \\ {g_1 \mu^{\epsilon_1}\over N} \mathcal{O}_1+ {g_2 \mu^{\epsilon_2}\over N} \mathcal{O}_2 
 + {g_3 \mu^{\epsilon_3}\over N} \mathcal{O}_3  + \mathcal{O}(\epsilon_i^2) ~, \label{potential}
\eea
where 
\bea
  \mathcal{M}_\phi(\beta) &=& 2 {g_1\mu^{\epsilon_1}\over N}\Big(1 +{2\over N}\Big) \langle \phi^2\rangle_\beta
  + {g_2\mu^{\epsilon_2}\over N} \langle \sigma^2\rangle_\beta ~,
  \non
   \mathcal{M}_\sigma(\beta) &=&  {g_2\mu^{\epsilon_2}\over N} \langle \phi^2\rangle_\beta 
   +6\,{g_3\mu^{\epsilon_3}\over N}\langle \sigma^2\rangle_\beta~.
   \label{M}
\eea
As shown in the Supplemental Material, thermal expectation values of the generalized free fields are given by
\be
\label{tev}
\langle \phi^2\rangle_\beta=N {2\,  \zeta(2\Delta_\phi)\over \beta^{2\Delta_\phi}},\quad 
\langle \sigma^2\rangle_\beta={2\,  \zeta(2\Delta_\sigma)\over \beta^{2\Delta_\sigma}}
 \ee
which follow straightforwardly from the thermal Green's function. Obviously, full $O(N)\times\mathbb{Z}_2$ symmetry is preserved by the minimum of the potential at zero temperature. 
However, the finite temperature effects might break it if $ \mathcal{M}_\phi<0$ or  $\mathcal{M}_\sigma<0$. 
If this happens, the higher order perturbative corrections cannot restore the symmetry, because the higher loop 
contributions to $\mathcal{M}_\phi, \mathcal{M}_\sigma$ are suppressed by higher powers of $\epsilon_i$, 
whereas the terms with higher powers of fundamental fields are subdominant in the vicinity of the origin. 
To establish symmetry breaking at finite temperature, it is therefore sufficient  to show that the model 
admits a fixed point where one of the masses  becomes negative.

To ensure stability of the model it is necessary to satisffy $g_1, g_3\geq 0$, while  $g_2$ could be negative provided 
 $4g_1g_3\geq g_2^2$. 
If all $g_i$ are positive, $\mathcal{M}_\phi, \mathcal{M}_\sigma$ are positive as well, the potential is minimized at the origin $\phi=\sigma=0$ 
and the symmetry is restored. The only scenario 
of symmetry breaking is therefore when  $g_2<0$. 
The RG flow \reef{RGflow} terminates at a weakly interacting IR fixed point in the vicinity of the original Gaussian theory. 
At the critical point at leading order in $\epsilon_i$ the couplings satisfy
\bea
\epsilon_1 g_1&=& {\pi^{d/2} \over N\Gamma\({d\over 2}\)} \big( C_{11}^1 g_1^2 + C_{22}^1 g_2^2   \big) ~,
\non
\epsilon_2 g_2&=& {\pi^{d/2} \over N\Gamma\({d\over 2}\)} \( 2 \, C^2_{12} \, g_1 g_2 + C^2_{22} g_2^2 
+ 2\,C^2_{23} g_2g_3\)~,
\non
 \epsilon_3 g_3&=& {\pi^{d/2} \over N\Gamma\({d\over 2}\)} \(C_{22}^3 \, g_2^2 +  C_{33}^3 \, g_3^2  \)~.
 \label{fp}
\eea
There is always a trivial fixed point with $g_1= {\Gamma\({d\over 2} \) \over 8 \pi^{d/2}} {N\over N+8} \epsilon_1$,  
$g_2=0$ and $g_3 = {N \Gamma\({d\over 2} \) \over 72\pi^{d/2}} \epsilon_3$.
 It represents two decoupled theories: the so-called Long Range Ising model \cite{Paulos:2015jfa} and its $O(N)$ generalization.
This fixed point was recently studied in, \eg \cite{Behan:2017dwr}.

\begin{figure}
\includegraphics[width=0.49\textwidth]{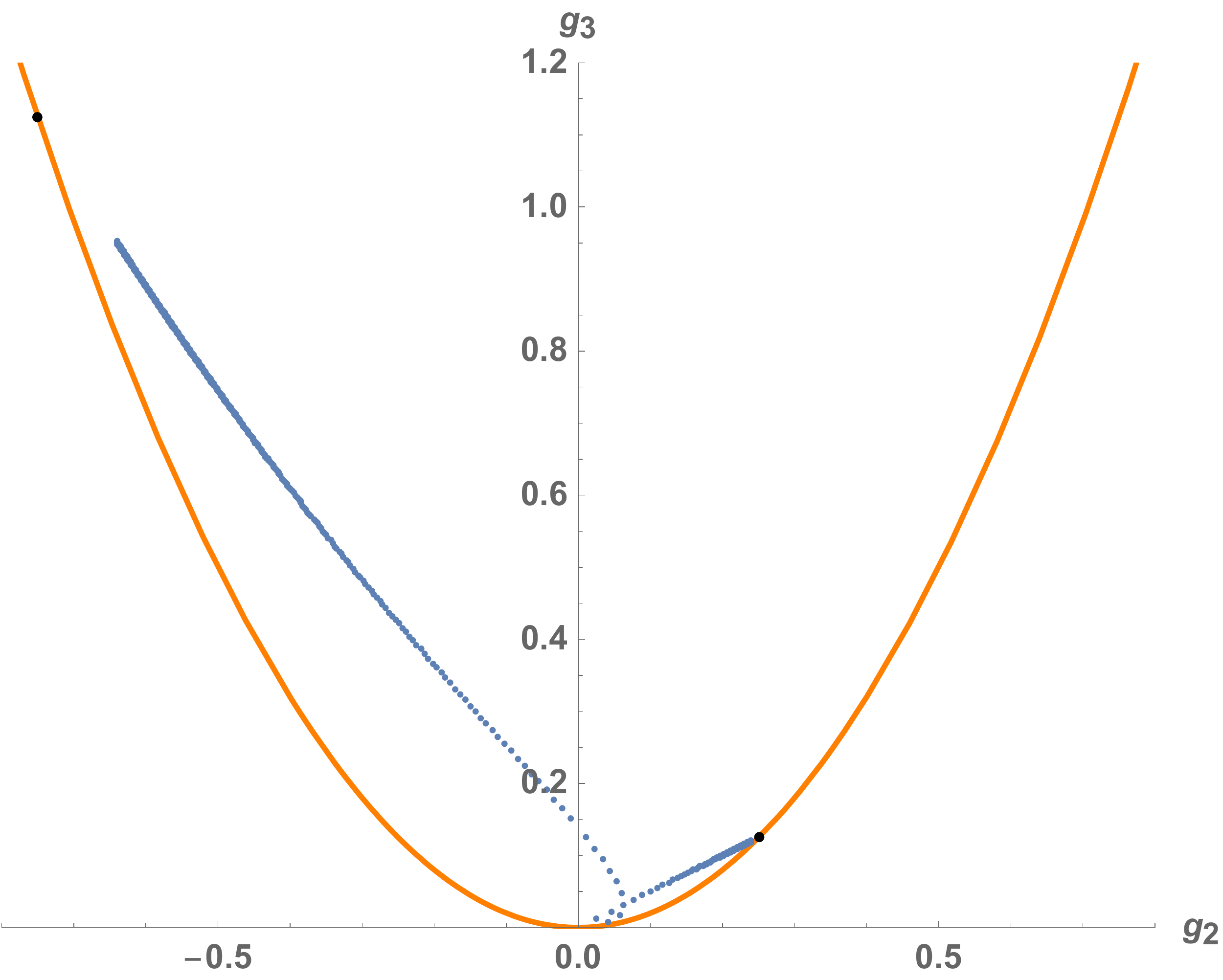}
\caption{Fixed points in the $g_2-g_3$ plane. Orange line shows a continuous family of solutions \eqref{family} emerging in the infinite $N$ approximation. Black 
dots are the asymptotic locations of fixed points  \eqref{oneoverN} in the limit of large $N$. Blue points are the solutions of \eqref{fp} for $N\leq 200$. }
\label{fig:parabola}
\end{figure}

To simplify the analysis and illustrate  the main idea in what follows we consider only a particular case of equal $\epsilon_i=\epsilon$.
The  case of non-equal $\epsilon_i$ is similar and also admits persistent symmetry breaking. It is also convenient to  rescale the couplings 
$g_i = \tilde g_i {\Gamma\({d\over 2} \) \over \pi^{d/2}} \epsilon_i$.
Before proceeding with the case of finite $N$ we take infinite $N$ limit. In this case the equations \eqref{fp} drastically simplify yielding the conformal manifold -- a one-parameter family of
fixed points 
\bea
 \tilde g_1= {1 \over 8} \,, \quad 
 \tilde g_3=2 \, \tilde g_2^{\, 2}, \label{family}
\eea
depicted in Fig.~\ref{fig:parabola}. 
For the negative branch $\tilde{g}_2=-2\sqrt{\tilde{g}_1\tilde{g}_3}$ the  effective potential degenerates into
\bea
 \label{eff_pot}
  V_\text{eff}(\phi, \sigma; \beta)= {\mu^\epsilon \over N} \( 2 \sqrt{g_1} \langle \phi^2\rangle_\beta\,  x+ x^2\) ~,\\
x=\sqrt{g_1} \phi^2 - \sqrt{g_3} \sigma^2.
\eea
The minimum is reached at  
\be
 \sigma^2 =\sqrt{{g_1\over g_3}} \( \phi^2+ \langle \phi^2\rangle_\beta\).
 \label{moduli}
\ee
This is a one-dimensional family of vacua
with  the non-zero expectation value of $\sigma$, signaling spontaneous symmetry breaking. For positive $g_2$ and $\langle \phi^2\rangle_\beta$ only $\sigma=0$ is admitted. 

From the discussion above it is clear negative $g_2$ is necessary for symmetry to be broken. 
Hence  the crucial question is if the  fixed point(s) with $g_2<0$ survive in the finite $N$ regime. 
Before proceeding with an arbitrary $N$ we employ $1/N$ expansion to find, in addition to \eqref{family},  the consistency condition 
\bea
\label{oneoverN}
4\tilde{g}_2^2+2\tilde{g}_2-{3\over 4}=0.
\eea
Hence at large but finite  $N$ the continuous family \eqref{family} collapses into two solutions, one with positive and one with negative $g_2$. 
Upon taking finite $N$ corrections into account  degeneracy $4g_1 g_3=g_2^2$ is lifted.
Minimization of the potential  \eqref{potential} yields
\be
  \left(\begin{array}{c}\phi^2 \\ \sigma^2 \end{array}\right) = {-N\mu^{-\epsilon} \over 4g_1g_3-g_2^2}
  \left(\begin{array}{cc}2g_3 & -g_2 \\-g_2 & 2g_1\end{array}\right) 
  \left(\begin{array}{c} \mathcal{M}_\phi \\ \mathcal{M}_\sigma\end{array}\right), 
    \label{vacuum}
\ee
provided resulting $\phi^2$ and $\sigma^2$ both are positive. Yet this is never the case for any solutions of \eqref{fp}.
The true minimum of the potential is therefore achieved either at $\phi^2=0$ or $\sigma^2=0$. For all solutions of \eqref{fp} $\mathcal{M}_\phi>0$ but 
$\mathcal{M}_\sigma$ become negative for the branch with negative $g_2$ and  $N>17$. In these cases the minimum is achieved at 
\bea
\left(\begin{array}{c}\phi^2 \\ \sigma^2 \end{array}\right) =    {-N\mu^{-\epsilon} \over 2 g_3}   \left(\begin{array}{c} 0 \\  \mathcal{M}_\sigma  \end{array}\right) ~.
\eea
Clearly, $\sigma^2$ is strictly positive, indicating symmetry breaking. 

\section{Conclusions}
To summarize, the model \eqref{action} with the choice $\epsilon_i=\epsilon \ll 1$ for finite $N$ has two different IR fixed points, 
and for  $N>10$ one of them has negative $g_2$.
The fixed points with $g_2<0$ and  $N>17$ exhibit  symmetry breaking at arbitrary non-zero temperature 
\be
 O(N)\times \mathbb{Z}_2 \quad \to \quad O(N)~.
\ee
The behavior for non-equal small $\epsilon_i$ is qualitatively similar. 

The model \eqref{action} admits a straightforward generalization to two Gaussian free fields in the vector representations of $O(N_1)$ and $O(N_2)$. 
The example in this note corresponds to $N_2=1$ case. Deforming this theory by weakly relevant quartic operators and following the 
same steps as above, one finds IR fixed points which exhibit spontaneous breaking of the global continuous symmetry at finite temperature. 
In $2+1$ dimensions it is therefore an example of persistent breaking of a continuous global symmetry, 
which bypasses the Coleman-Hohenberg-Mermin-Wagner theorem \cite{Mermin:1966fe,PhysRev.158.383,Coleman:1973ci} by virtue of being a CFT with long-range interactions. We will discuss this case in detail in \cite{followup}. Our non-local $O(N)\times \Z_2$ model is similar to that one studied in \cite{Chai:2020zgq,Chai:2020onq}. It would be also interesting to explore if they are related in the  large $N$ limit.

Our findings clearly show persistent breaking is possible in the UV-complete  yet non-local models. This raises the question if non-locality is truly necessary, i.e.~if there could be UV-complete unitary {\it local} field theoretic models in $d=2+1$ exhibiting persistent breaking of  discrete symmetries. It is an important open question to construct such an example or rule out this possibility.\\

{\bf Acknowledgements}  We thank Alex Avdoshkin, Dean Carmi, Soumyadeep Chaudhuri, Changha Choi, Joshua Feinberg, Mikhail Goykhman, Eliezer Rabinovici, Ritam Sinha and especially Zohar Komargodski for helpful discussions and correspondence. 
This work is partially supported by the Binational Science Foundation (grant No. 2016186). 
AD is grateful to Weizmann Institute of Science for hospitality and acknowledges  sabbatical support of the Schwartz/Reisman Institute 
for Theoretical Physics.
NC and MS  are grateful to the Israeli Science Foundation Center of Excellence (grant No. 2289/18) and the Quantum Universe I-CORE program of the Israel Planning and Budgeting Committee (grant No. 1937/12) for continuous support of our research. NC is grateful for the support from the Yuri Milner scholarship.

\appendix
\section{ Supplemental Material}
The action defines a conformal model with the canonically normlized two-point function of fundamental fields $\vec\phi$ and 
$\sigma$,
\bea
\langle \phi_a \phi_b \rangle={\delta_{ab} \over |x_{12}|^{2\Delta_\phi}},\quad
\langle \sigma \sigma \rangle={1\over |x_{12}|^{2\Delta_\sigma}}.
\eea
The two-point functions  of the operators $O_i$ \eqref{Oi}
\bea
\langle\mathcal{O}_i \mathcal{O}_j \rangle=\delta_{ij}{N_{i}\over |x|^{2\Delta_i}}
\eea 
where 
\bea
 N_1 = {8 N^2\Big(1+{2\over N}\Big)} , \quad N_2= {4 N} ,\quad 
N_3= {24}~.
 \label{2p}
\eea
Similarly, the three-point functions 
\bea
\nonumber
\langle\mathcal{O}_i \mathcal{O}_j \mathcal{O}_k\rangle&=&{C_{ij}^k N_k \over
|x_{12}|^{\Delta-2\Delta_k} 
|x_{23}|^{\Delta-2\Delta_i} 
|x_{13}|^{\Delta-2\Delta_j}},\\
\Delta&=&\Delta_i+\Delta_j+\Delta_k,
\eea
are fixed by the OPE coefficients (we list only non-zero ones)
\bea
 &&C_{11}^1= 8 \,  (N+8) ~, ~ C_{22}^1=2~, ~ C^2_{12}=4\(N+{2}\)
 ~,  
 \non
 && C^2_{22}=16~, ~ C^2_{23}=12~, ~ C_{22}^3=2N ~, ~ C_{33}^3=72 ~.
\eea
They are related by
\bea
C_{ij}^k =C_{ik}^j N_j/N_k.
\eea

At finite temperature two-point function takes the form 
\bea
\langle \phi_a \phi_b \rangle=\sum_{m=-\infty}^\infty {\delta_{ab}\over \[ (\tau + m \beta)^2 + \vec x^2\]^{\Delta_\phi}},
\eea
and similarly for $\sigma$. From here one trivially finds \eqref{tev}.

\bibliographystyle{unsrtnat}
\bibliography{Letter}

\end{document}